\documentclass	[prb,twocolumn,showpacs]{revtex4-1}%
\usepackage{graphicx}%
\usepackage{dcolumn}%
\usepackage{bm}
\usepackage{amsmath}
\usepackage{epstopdf}
\usepackage{color}
\usepackage{lipsum}
\usepackage{float}

\begin{document}

\title{Azimuthal Spin Wave Excitations in Magnetic Nanodots over the Soliton Background: Vortex, Bloch and N\'eel-like skyrmions}
\author{M.~Mruczkiewicz$^{1}$ } \email{mmruczkiewicz@gmail.com}
\author{P.~Gruszecki$^{2}$ } 
\author{M.~Krawczyk$^{2}$ } 
\author{K. Y. Guslienko$^{3,4}$ } 
\affiliation{$^{1}$Institute of Electrical Engineering, Slovak Academy of Sciences, Dubravska cesta 9, 841 04 Bratislava, Slovakia\\
$^{2}$Faculty of Physics, Adam Mickiewicz University in Poznan, Umultowska 85, Pozna\'{n}, 61-614, Poland \\
$^{3}$Depto. Fisica de Materiales, Universidad del Pais Vasco, UPV/EHU, 20018 San Sebastian, Spain \\
$^{4}$IKERBASQUE, the Basque Foundation for Science, 48013 Bilbao, Spain}

\date{\today}

\begin{abstract}

We study azimuthal spin-wave (SW) excitations in a circular ferromagnetic nanodot in different inhomogeneous, topologically non-trivial magnetization states, specifically, vortex, Bloch-type skyrmion and N\'eel-type skyrmion states. Continuous mapping of the SW spectrum between these states is realized with gradual change of the out-of-plane magnetic anisotropy and Dzyaloshinskii-Moriya exchange interaction (DMI). Our study shows lifting of the SW frequencies degeneracy and change in systematics of the frequency levels. The change is induced by a geometrical Berry phase, that is present for the dot-edge localized SWs in a vortex state and vanishes in skyrmion states. Furthermore, channeling of the azimuthal SWs localized at the skyrmion edge is present and induces large frequency splitting. This is attributed to DMI induced nonreciprocity, while coupling of the breathing and gyrotropic modes is related to soliton motion. Finally, an efficient coupling of the dynamic magnetization in the skyrmion state to uniform magnetic field in nanodots with non-circular symmetry is shown.

\end{abstract}

\maketitle

\section{Introduction}

The patterned magnetic nanostructures are considered as novel devices for information processing and storage \cite{chui19magnetic,5} that can go beyond limitations of the semiconductor technology. \cite{khitun2012multi} In particular, ultrathin magnetic structures with an interface induced Dzyaloshinskii-Moriya interaction (DMI) have become subject of intensive research in that direction. \cite{boulle2016room,moreau2016additive}
The DMI interaction energy is minimized when neighboring spins are perpendicular, with specific sense of rotation. In this sense, DMI interaction introduces chirality (handedness). If the DMI
contribution to the magnetic energy is sufficiently large, the magnetization configuration in the ground state is inhomogeneous (spiral or skyrmion states). Such states are promising for applications. For example, skyrmion state is considered as a stable nm-size topological magnetic soliton, manipulated applying low energy external stimuli. 

Although the DMI facilitates stabilization of inhomogeneous magnetization states, there is a broad range of the parameters (e.g., low DMI strength, external magnetic field) for which the single domain (SD) configuration is still a ground state. \cite{garcia2014nonreciprocal,fert2013skyrmions}
Even though the presence of DMI does not result in change of the static magnetization  configuration, its
influence on magnetization dynamics is essential. It has been shown that DMI can influence frequency of spin wave (SW) propagating at certain magnetization arrangement (i.e., perpendicular to the magnetization). \cite{udvardi2009chiral,moon2013spin,cortes2013influence}
DMI alters the symmetry of the SW dispersion relation introducing nonreciprocity
($f(-k)\neq f(k)$). The effect is commonly used to extract the DMI strength from the measured SW spectra.

The nonreciprocal SW propagation induced by DMI may also exists
in confined geometries. We have shown with numerical calculations
and simulations that the effect of SW nonreciprocity is present
in magnetic nanostripes.\cite{mruczkiewicz2016influence} The spectrum of standing SWs in an isolated stripe is modified and significant increase of quantized modes coupling to the spatially uniform microwave field is expected. Since a
decrease of time required for measurements of the DMI strength is in search,\cite{hrabec2017making}
an experimental confirmation of that effect would open an alternative way to Brillouin Light Scattering (BLS) for the DMI strength estimation.

The SW excitation modes of different symmetry have been studied extensively in the single domain stripes,\cite{guslienko2002effective,bayer2003spin}
disks, \cite{giovannini2004spin,buess2004fourier,park2005interactions,buess2005excitations,zhu2005broadband,zivieri2005theory,zaspel2005excitations,lupo2015size, kakazei2004spin}, 
ellipses,\cite{gubbiotti2005spin,demidov2010nonlinear,nembach2011effects}
rings,\cite{podbielski2006spin,neudecker2006spatially,gubbiotti2006splitting,giesen2007mode,schultheiss2008observation}
nanotubes\cite{otalora2017asymmetric}. Recently also the influence of DMI on the magnetization stable state\cite{dejong2017domain} and dynamic excitations have been investigated.\cite{moon2013spin,garcia2014nonreciprocal}
It has been demonstrated that the SW channeling within a domain wall is possible
and could lead to reconfigurable channels and controllable signal propagation.\cite{wagner2016magnetic,lara2017information}     Nonreciprocity and DMI introduce another degrees of freedom  in controlling SWs.\cite{garcia2015narrow}
The skyrmion dynamics have already been studied in circular geometry\cite{kim2014breathing,gareeva2016magnetic,mruczkiewicz2016collective,beg2017dynamics,guslienko2017gyrotropic,mruczkiewicz2017spin,liu2017shape, li2017creation,garst2017collective,kim2016coupled}  but
with limited consideration of the quantization of azimuthal SW modes. Since the magnetic configuration can favour the nonreciprocity in many cases, it is expected that a significant influence on azimuthal SW spectrum can be induced
by the DMI.

\begin{figure}[!ht]
\includegraphics[width=0.35\textwidth]{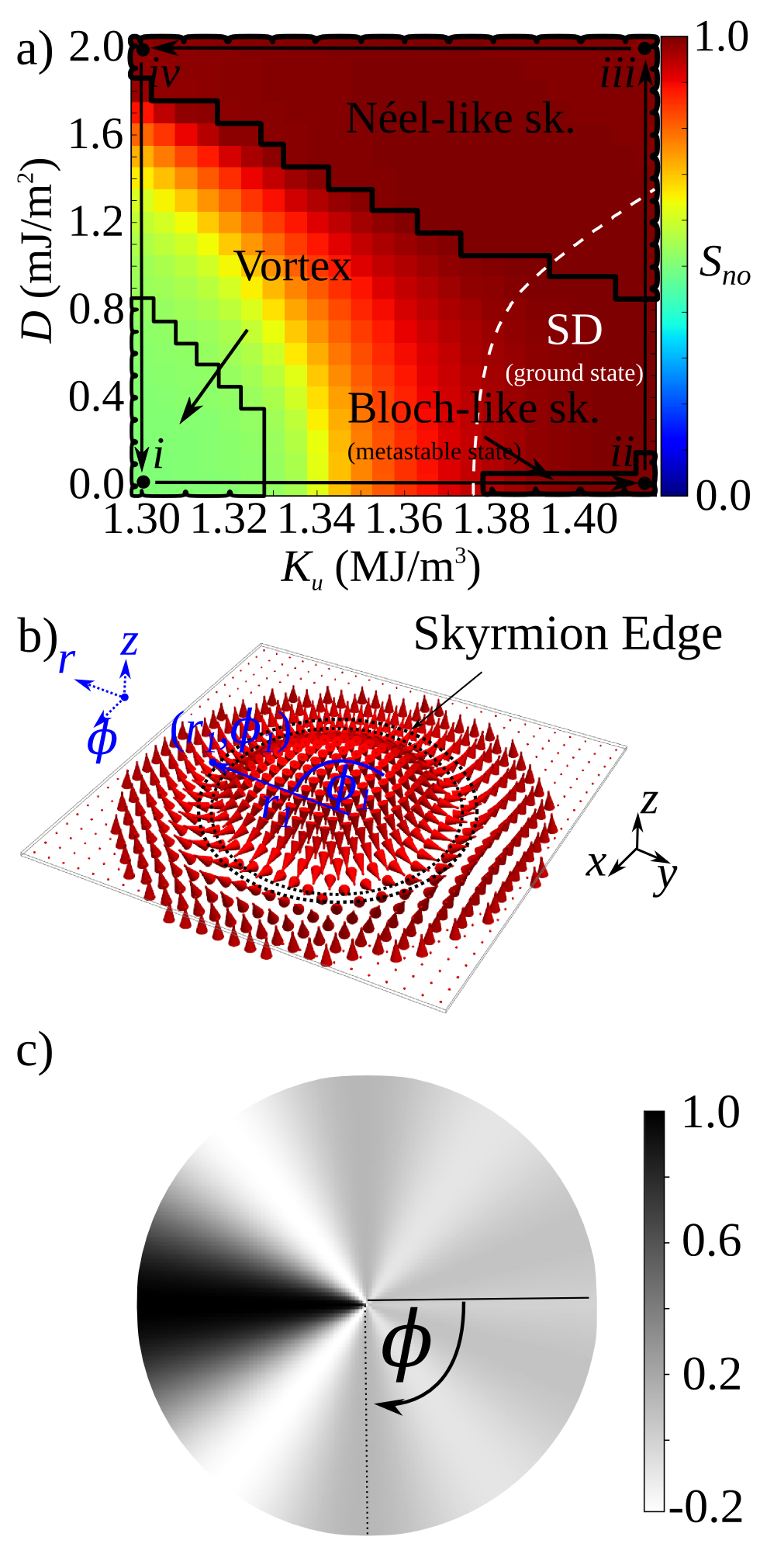} \caption{(Color online) a) The skyrmion number $S_{\mathrm{no}}$ in a circular ferromagnetic dot as a function of magnetic anisotropy ($K_{\mathrm{u}}$) and DMI strength, exhibiting continuous transitions between four magnetization states (reprinted with permission from [\onlinecite{mruczkiewicz2017spin}]. Copyright (2017) by the American Physical Society, https://doi.org/10.1103/PhysRevB.95.094414). b) N\'eel-like skyrmion in nanodot with strong perpendicular magnetic anisotropy (large value of the quality factor $Q=\frac{2K_{\mathrm{u}}}{\mu_{0}M_{\mathrm{s}}^{2}}$).
c) The distribution of the amplitude of the excitation magnetic field, $\bf B_{\mathrm{ext}}= I$sinc$(k_{\phi}(\phi-\pi))$sinc$(2\pi f_{\mathrm{max}}t)(1,\,0,\,1)$.
}
\label{mathematica} 
\end{figure}

In this paper we study influence of the DMI on the azimuthal SW localization and frequency over the skyrmion magnetization background in a circular nanodot with reference to the vortex state. We show, that the topological Berry phase \cite{dugaev2005berry} is responsible for lifting degeneracy of a dot-edge localized SW and change in systematics of the eigenfrequencies. Moreover, the DMI inducing nonreciprocity gives rise to splitting of the frequencies of SWs with the opposite directions of propagation and leads to creation of additional low frequency quantized modes. We also demonstrate that investigated magnetization configurations can be efficiently excited by the uniform microwave field in nanoelements with non-circular symmetry. The results can serve to control magnetization dynamics in  patterned nanostructures with DMI, SW channeling and control of information processing by SWs. 

\section{Model}

\label{Sec:Model} The physical system we consider is a thin circular ferromagnetic
dot of the thickness $t$ and radius $R$. To find the dot spin excitation
spectrum the finite difference time domain (FDTD) micromagnetic simulations
were performed using mumax$^{3}$ code.\cite{4899186} We start from the Landau-Lifshitz equation for magnetization ${\bf M}$ dynamics in
which the magnetization time derivative $\frac{\partial{\bf M}({\bf r},t)}{\partial t}$
is defined as the torque ${\bf \tau}$ that can be expressed in the
following form:
\begin{equation}
{\bf \tau}=\left|\gamma\right|\frac{1}{1+\alpha^{2}}\left({\bf M}\times{\bf B_{\mathrm{eff}}}+\alpha\left({\bf M}\times\left({\bf M}\times{\bf B_{\mathrm{eff}}}\right)\right)\right),
\end{equation}
where $\gamma$ is the gyromagnetic ratio, $\alpha$ is a dimensionless
damping parameter, and ${\bf B_{\mathrm{eff}}}$ is the effective
magnetic field, which includes the external magnetic field ${\mathbf{B}_{\text{ext}}}$,
the magnetostatic field ${\bf B_{\mathrm{m}}}$,
the isotropic Heisenberg exchange field ${\bf B_{\mathrm{ex}}}$
(being proportional to the exchange stiffness constant $A$), the interface Dzyaloshinskii-Moriya
exchange field ${\bf B_{\mathrm{DM}}}$, and the uniaxial magnetocrystalline
anisotropy field ${\bf B_{\mathrm{an}}}$ (being proportional to
the uniaxial anisotropy constant $K_{\mathrm{u}}$): 
\begin{equation}
{\bf B_{\mathrm{eff}}} = {\mathbf{B}_{\text{ext}}}+{\bf B_{\mathrm{m}}}+{\bf B_{\mathrm{ex}}}+{\bf B_{\mathrm{DM}}}+{\bf B_{\mathrm{an}}}.
\end{equation}

The skyrmions are stabilized in the dot due to an interplay of the
isotropic exchange, DMI, uniaxial out-of-plane magnetic anisotropy,
and magnetostatic energies assuming zero bias magnetic field. According to Ref. [\onlinecite{bogdanov2001chiral}] the
DMI is implemented as an effective magnetic field 
\begin{equation}
\mathbf{B}_{\mathrm{DM}}=\frac{2D}{M_{\mathrm{s}}}\left(\frac{\partial m_{z}}{\partial x},\frac{\partial m_{z}}{\partial y},-\frac{\partial m_{x}}{\partial x}-\frac{\partial m_{y}}{\partial y}\right)
\end{equation}
and give rise to the magnetic energy density 
\begin{equation}
\varepsilon=D\left(m_{z}\left(\nabla\cdot\mathbf{m}\right)-\left(\mathbf{m}\cdot\nabla\right)m_{z}\right),
\end{equation}
where $\mathbf{m}=\mathbf{M}/M_{\mathrm{s}}$ is the reduced magnetization
vector, $M_{\mathrm{s}}$ is the saturation magnetization and $D$
is the Dzyaloshinskii-Moriya interface exchange interaction constant.


The simulations consist of the following steps. The initial magnetization configuration 
was assumed in the form of the Bloch skyrmion. This initial state
was then relaxed in order to minimize the total magnetic energy of the system.
During that process magnetic configuration transforms into either
vortex, Bloch skyrmion or N\'eel skyrmion state, depending on the chosen magnetic 
parameters ($K_{\mathrm{u}}$, $D$). In majority of the cases, the
initial Bloch-like skyrmion state relaxes to one of the ground states. However,
for some combinations of the dot magnetic parameters, the
Bloch-like skyrmion becomes metastable state and does not relax to
the ground state.  This region is indicated by the white dashed line
in Fig. \ref{mathematica} (a). In this case a single domain (SD) state
can be used as an initial magnetization state in order to obtain the ground state.

\begin{figure}[!ht]
\includegraphics[width=0.45\textwidth]{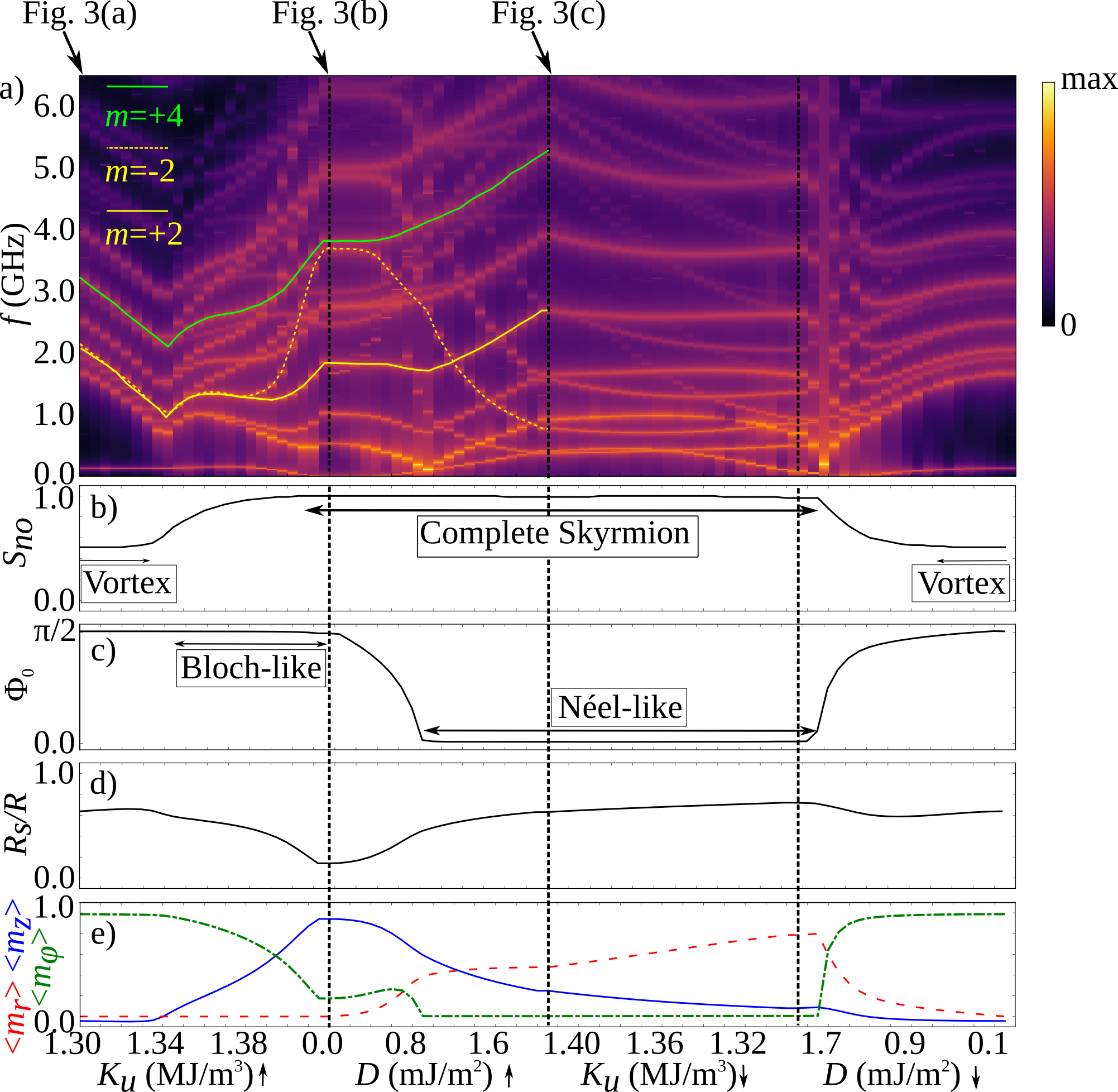} \caption{(Color online) (a) The soliton spin wave frequencies
excited with nonuniform magnetic field are plotted along the path presented
in Fig. \ref{mathematica} (a) between the four (i-iv) inhomogeneous
magnetization configurations.  The static properties
of the solitons are presented in the panels (b) skyrmion number $S_{\mathrm{no}}$,
(c) averaged skyrmion phase, $\Phi_{0}$, (d) skyrmion size $R_{s}/R$ and
(e) the averaged magnetization components, $<m_{r}>$, $<m_{\phi}>$
and $<m_{z}>$ with red dashed, green dot-dashed and blue continuous
lines, respectively.}
\label{disp} 
\end{figure}

The stable magnetization configurations were excited with low
amplitude variable magnetic field having a time dependence represented
by the function $\mathrm{sinc}\left(2\pi f_{\mathrm{max}}\left(t-t_{0}\right)\right)$
with the cut-off frequency $f_{\mathrm{max}}=10$ GHz and
$t_{0}=1$ ns.\footnote{The cell size below 1 nm was used, $\frac{250\times10^{-9}}{256}$.
The maximum amplitude of the $\mathrm{sinc}$ signal was $I=0.5$
mT.} This value of the cut-off frequency was chosen because we
are interested to map the low frequency part of the SW
excitation spectra related to the different magnetization
states of the dot. 

In order to excite and detect higher order azimuthal modes (these with azimuthal mode index $\vert m \vert>1$,  defined later),
in addition to time modulation, a spatial modulation (dependent on the polar angle $\phi$) of the driving magnetic field was introduced, $\mathrm{sinc}(k_{\phi}(\phi-\pi))$,
with $k_{\phi}=2.5$. Therefore, several azimuthal SW with $\vert m \vert>1$ could be efficiently coupled to the driving field (see Fig. \ref{mathematica}(c) for the distribution of the field in the circular dot). Thus, the  magnetic field with amplitude $I$ used for excitation has the following form: 
\begin{equation}
\begin{split}
\mathbf{B}_{\mathrm{ext}}=[ I\mathrm{sinc}\left(k_{\phi}(\phi-\pi)\right)\mathrm{sinc}\left(2\pi  f_{\mathrm{max}}\left(t-t_{0}\right)\right), 0,  \\ I\mathrm{sinc}\left(k_{\phi}(\phi-\pi)\right)\mathrm{sinc}\left(2\pi  f_{\mathrm{max}}\left(t- t_{0}\right)\right) ].
\end{split}
\end{equation}

The space and time dependent magnetization components acquired after
the field excitation were transformed to the frequency domain
(Fourier transform \footnote{The rectangular window function was used in the Fourier transform
with 50 ns width, starting 4 ns after the peak of the sinc signal,
providing 20 MHz frequency resolution, $\Delta f=20$ MHz}) to obtain the power spectral density (PSD), SW eigenfrequencies and spatial distribution of the dynamical components of the selected eigen oscillations of the magnetization vector $\mathbf{m}$. \cite{mruczkiewicz2016collective}

We consider linear dynamics over the soliton magnetization background, $\Theta=\Theta_{s}+ \vartheta$, $\Phi=\Phi_{s}+ \psi$, where the static skyrmion magnetization spherical angles are $\Theta_{s},\Phi_{s}= \Phi_{0}+ \phi$, and $\vartheta, \psi$ are the SW angles. Two dynamic magnetization components in the local coordinate system with $Oz^{\prime}$ axis directed along the skyrmion static magnetization, $\mathbf{m}(\Theta_{s},\Phi_{s})$ can be represented as:
\begin{equation}
\begin{split}
{\bf \delta m}=  [ \vartheta, sin(\Theta_{s}) \psi, 0]=\\
=[a_{\mathfrak{n}}(r)\cos(m\phi-\omega t), b_{\mathfrak{n}}(r)\sin(m\phi-\omega t), 0],
\end{split}
\end{equation}
where $a_{\mathfrak{n}}$, $b_{\mathfrak{n}}$ are the SW mode radial profiles, $\mathfrak{n}$ is the number of nodes along the radial direction (radial mode index), $m$ is the azimuthal mode
index, and $\phi$ is a polar coordinate of the cylindrical system. In the folowing we consider the simplest radial mode with $n=0$. For radially symmetric static magnetization configurations $\Theta_{s}= \Theta_{0}(r)$ and $\Phi_{0} =\Phi_{0}(r)$ \cite{7061384}. The type of the skyrmion (Bloch-like or N\'eel-like) are distinguished with a function $\Phi_{0}$ representing the skyrmion phase. It takes values $ \pm  \frac{\pi}{2}$ for the magnetic vortex or complete Bloch-like skyrmion (the in-plane magnetization is aligned along the azimuthal direction everywhere in the dot including the vortex/skyrmion edge) and $0, \pi$ for the complete N\'eel-like skyrmion (the in-plane component of the magnetization is along the radial direction).

The soliton dynamical magnetization
components in the laboratory frame (cylindrical coordinate system) $\delta m_{r}$, $\delta m_{\phi}$, $\delta m_{z}$ can be represented as linear combination of these two magnetization components given by Eq. (6) 
involving the skyrmion static magnetization angles $\Theta_{s}$ and $\Phi_{s}$.\cite{guslienko2016gauge} Accordingly,  the azimuthal index $m$ determine the number of nodes in the $\delta m_{r}$, $\delta m_{\phi}$ or $\delta m_{z}$ distributions along the polar angle $\phi$. However, for excitation of the azimuthal modes an in-plane variable field should be applied along $Ox$ or $Oy$ directions. Therfore, we rewrite the dynanamic magnetization components in the laboratory coordinate system $xyz$, 

\begin{equation}
\begin{split}
\delta m_{x} + i  \delta m_{y} =(\delta m_{r} + i  \delta m_{\phi})exp(i \phi)= \\
=(\vartheta cos(\Theta_{s}) +i \psi sin( \Theta_{s})) exp(i(\Phi_{0} + \phi)),\\
\delta m_{z}= -sin(\Theta_{s}) \vartheta.
\end{split}
\end{equation}

For the cartesian components $\delta m_{x}, \delta m_{y}$ the azimuthal mode index $m$ is shifted to $m+1$ in the inner skyrmion area $r<R_{s}$, and to $m-1$ in the outer skyrmion area  $r>R_{s}$, if the skyrmion polarization $p>0$. The index change is opposite for the case $p<0$. The cartesian magnetization components in the local coordinate system $x^{\prime}, y^{\prime}, z^{\prime}$ and the laboratory system $xyz$ are connected via the rotation matrix $\mathcal{R} $ parametrized by the Euler angles as $\delta \mathbf{m} = \mathcal{R}(\Theta_{s}, \Phi_{s}, 0) \delta \mathbf{m}^{\prime}$.  

Throughout the paper we use the following material parameters of an ultrathin magnetic circular dot of the radius $R=125$ nm and thickness $t=1.4$ nm: saturation magnetization $M_{\text{s}}=1.5\times10^{6}$
A/m, exchange stiffness constant $A=3.1\times10^{-11}$
J/m, DMI constant $D$ varies from $0$ to $3.8\times10^{-3}$ J/m$^{2}$,
and out-of-plane magnetic anisotropy constant $K_{\mathrm{u}}$ varies
from $1.30\times10^{6}$ J/m$^{3}$ to $1.415\times10^{6}$ J/m$^{3}$.
This set of parameters corresponds to the ultrathin layers of CoFeB-MgO.\cite{nakatani2016electric}
The material quality factor $Q=\frac{2K_{\mathrm{u}}}{\mu_{0}M_{\mathrm{s}}^{2}}$
varies from 0.92 to 1.0. The Gilbert damping
parameter taken into account in the FDTD simulations is $\alpha=0.01$.
It is close to the damping value of an ultrathin CoFeB film.\cite{Natarajarathinam12,Yu12} The present study is also extended to account for higher interface DMI values, bulk DMI and different dot sizes.

\section{Results and Discussion}

\begin{figure}[!ht]
\includegraphics[width=0.49\textwidth]{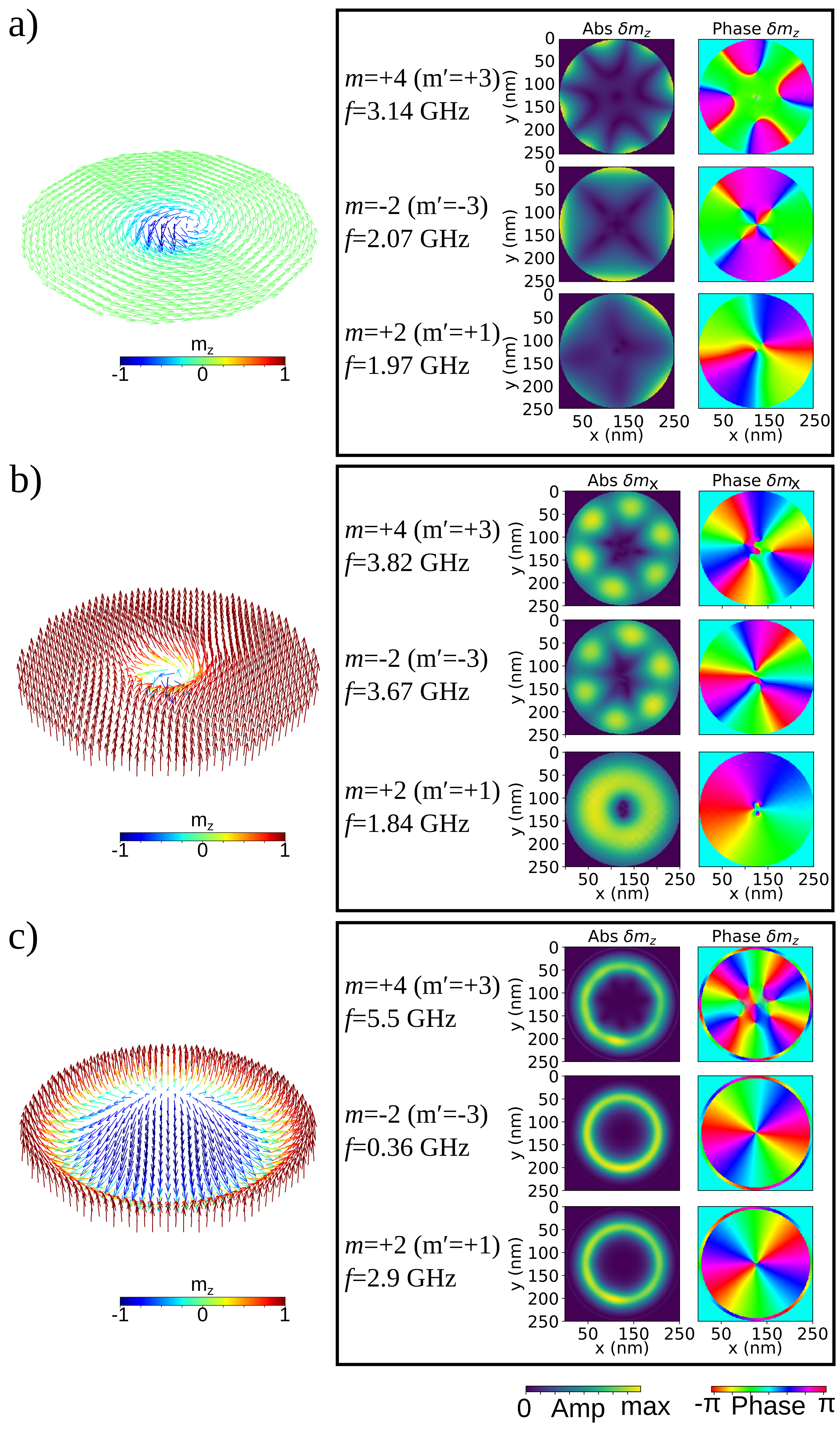} \caption{(Color online) The spatial distribution of the static magnetization (left column) and SW eigenmodes (right column) (amplitude and phase for $x$ or $z$ components of the dynamical magnetizations) for three different modes corresponding to the following azimuthal mode indices $m=+4, -2$ and $+2$. In brackets the numbers of nodes in the dynamic magnetization components, $m_{x}$ and $m_{y}$, are listed. The magnetic parameters are: a) $K_{\mathrm{u}}=1.30$
MJ/m$^{3}$, $D=0$, vortex (i); b)$K_{\mathrm{u}}=1.415$
MJ/m$^{3}$, $D=0$, Bloch-like skyrmion (ii) and c) $K_{\mathrm{u}}=1.415$ MJ/m$^{3}$, $D=2.0$ mJ/m$^{2}$, N\'eel-like skyrmion (iii).}
\label{profiles} 
\end{figure}

Figure~\ref{mathematica} (a) presents the skyrmion number as a function of the DMI strength and uniaxial anisotropy constant. It allows us to choose the path for study of the soliton dynamical excitations, where continuous transitions occur: between the vortex and Bloch-like skyrmion along the path (i) $\rightarrow$ (ii), between the Bloch-like skyrmion and N\'eel-like skyrmion with high $Q$ along the path (ii) $\rightarrow$ (iii), between
high and low $Q$ N\'eel-like skyrmions along the path (iii) $\rightarrow$ (iv), and finally between the low $Q$ N\'eel-like skyrmion and vortex state along the path (iv) $\rightarrow$ (i). These paths are indicated in Fig. \ref{mathematica} (a) with black straight arrows. In Fig. \ref{mathematica} (b) a high $Q$ N\'eel-like skyrmion is presented with indicated ring where the $z$-component of magnetization is approximately zero ($m_{z} \approx 0$), skyrmion edge. The magnetization states corresponding to points (i) - (iv) are presented in Ref. [\onlinecite{mruczkiewicz2017spin}].

\begin{figure*}
\centering
\includegraphics[width=0.9\textwidth]{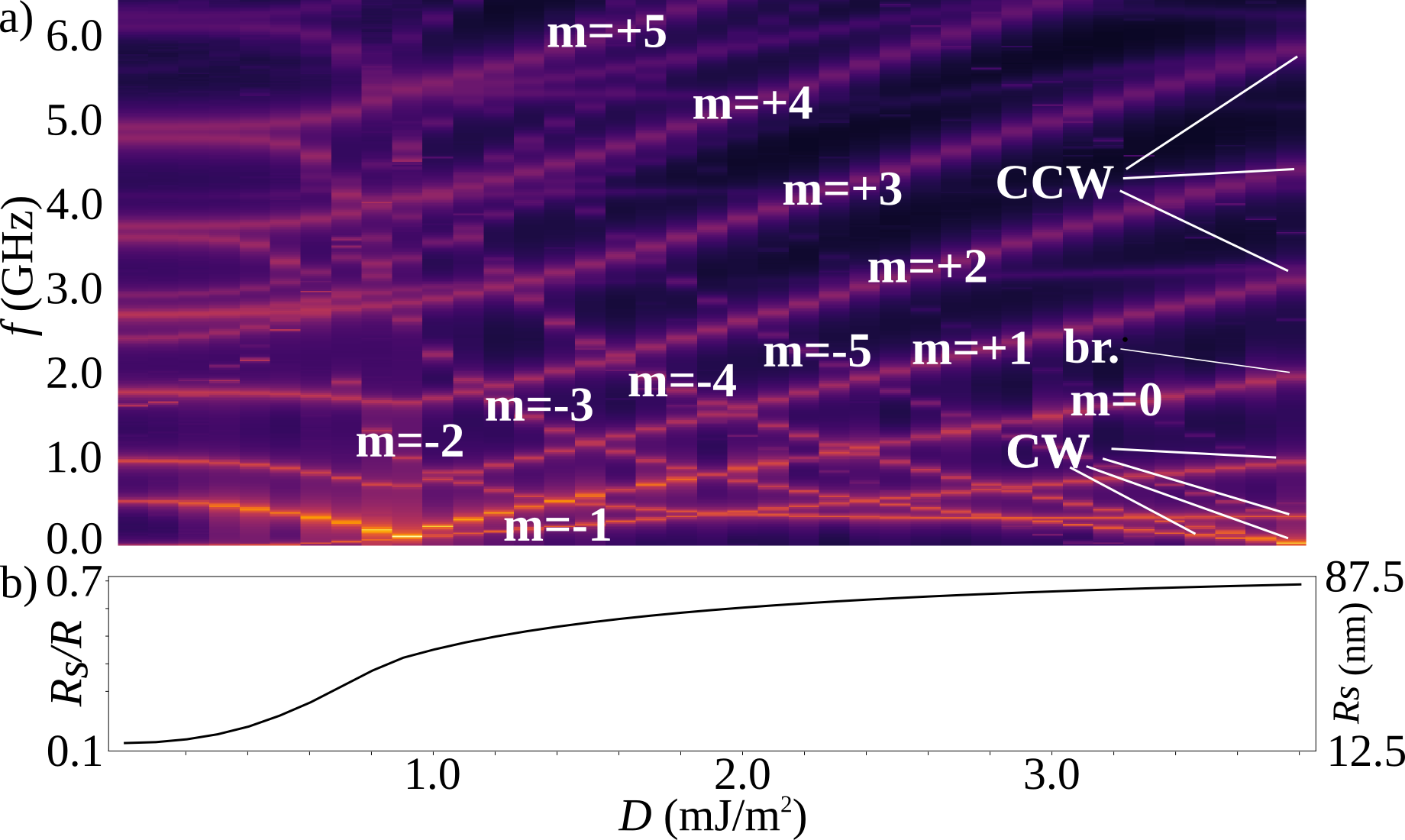} \caption{(Color online) a) Spectrum of the azimuthal SW modes at fixed value of $K_{\mathrm{u}}=1.415\times10^{6}$ J/m$^{3}$ as function of the DMI strength, $D=0-3.8 \times10^{-3}$ J/m$^{2}$. Frequency splitting of CW and CCW azimuthal modes is pronounced at high DMI, $D$, values. b) Skyrmion size dependence on $D$. }
\label{Fig4} 
\end{figure*}


The calculated frequencies of low-lying spin excitations along the defined above paths are presented in Fig. \ref{disp} (a). Numerous SW excitations are observed as compared to the modes excited with uniform magnetic field, presented in Ref. [\onlinecite{mruczkiewicz2017spin}]. The parameters characterizing the static skyrmions are plotted as a function of the dot magnetic parameters in Figs. \ref{disp} (b-e). The regions where magnetization configuration is a
complete vortex, Bloch-like, N\'eel-like skyrmion or an intermediate state can be clearly differentiated and are indicated in Fig.~\ref{disp}
(b), (c)  and also in
Fig.~\ref{mathematica}(a). Fig.~\ref{disp}(e) demonstrates the
second-order phase transitions, i.e., continuous transitions of the average magnetization components (the perpendicular $\langle m_{z} \rangle$, radial
$\langle m_{r} \rangle$ and azimuthal $\langle m_{\phi} \rangle$ components) along the paths.
We consider the full spin excitation spectra changing the magnetic parameters along the paths (i)$\rightarrow$(ii)$\rightarrow$(iii)$\rightarrow$(iv),
to make a complete comparison of the spectra with ones calculated in Ref. [\onlinecite{mruczkiewicz2017spin}].
However, the most of the discussion refers to the left half of
the spectrum with increasing the values of the anisotropy and DMI (the path (i)$\rightarrow$(ii)$\rightarrow$(iii)). 

In the dynamical simulations, there are spin eigenmodes keeping the radial symmetry of the soliton static state (radially symmetric or
breathing modes with $m=0$) and the eigenmodes which break this radial symmetry (azimuthal modes rotating in the clockwise (CW) and counter-clockwise (CCW) directions).
The radially symmetric modes have no net in-plane magnetization,
and therefore, can be excited only by the out-of-plane variable magnetic
field. \cite{mruczkiewicz2016collective}  The azimuthal modes with the indices $\vert m \vert=1$,
including gyrotropic one, can be excited by an uniform in-plane variable
magnetic field, whereas the modes with higher azimuthal indices ($\vert m \vert>1$)
cannot be excited with an uniform field due to the symmetry of their microwave magnetization distribution. The nonuniform excitation field described in Sec. II was used in order to excite the higher order azimuthal SW modes.

As presented in Fig. \ref{disp}(a), a continuous transition of the azimuthal SW mode frequencies is found. All azimuthal SW excitations over the vortex background have counterpart excitations in the skyrmion states. The first striking observation is lifting of frequency degeneracy of the modes with the indices $m=\pm \vert m\vert$, $\vert m\vert>1$.
In the vortex states theses modes are almost completely degenerated (splitting of their frequencies is negligibly small), and the degeneracy is defined by the azimuthal number $m$, see Fig. \ref{profiles}(a)
for distribution of the dynamical magnetization $z$-component $\delta m_{z}$. In the case of mixed states or N\'eel
and Bloch skyrmion states, degeneracy of these modes is lifted (indicated by continuous yellow and dashed yellow lines in Fig. \ref{disp}(a) above $K_{\mathrm{u}}=1.38\times10^{6}$
J/m$^{3}$ for m=+2 and -2, respectively). However, another degeneracy of the modes (change in systematics of the eigenfrequencies) is visible for the Bloch skyrmion (indicated by continuous green and dashed yellow lines in Fig. \ref{disp}(a) at $K_{\mathrm{u}}=1.415\times10^{6}$ J/m$^{3}$, for $m$=+4 and -2, respectively).

The reason for the degeneracy lifting for mixed states and different pairing of modes in the vortex/Bloch states is due to different magnetic configurations and regions where localization of the SW amplitude is strongest. In the case of vortex, azimuthal SWs with $\vert m \vert>1$ are localized near the dot edge, where in-plane magnetization is dominating and it curls around the center of the dot, possessing a cylindrical symmetry. With increasing the anisotropy, the static vortex state is transformed to a Bloch skyrmion, the out-of-plane magnetization component gradually increases. However, due to small size of the skyrmion, SW is still localized at the dot edge (see Fig. \ref{profiles}(b) for the  $x$-component of dynamical magnetization, $\delta m_{x}$), where the static magnetization is perpendicular to the dot plane. The number of azimuthal nodes of degenerated states in the Bloch-skyrmion states differs by $\pm1$, accordingly to the mode numbering defined in Sec. II. 

The gradual increase of anisotropy constant $K_{u}$ leads to the continuous transformation of the vortex to the Bloch-like skyrmion and the mode localized in the area of curling in-plane magnetization with cylindrical symmetry transforms to the mode localized in the area of out-of-plane magnetization. In that sense, the magnetization state that support azimuthal SW in transition between the vortex and Bloch-skyrmion is similar to SW propagating in a thin ring, as presented in  Fig. 5 in Ref. [\onlinecite{dugaev2005berry}]. There, the magnetic configuration in the ring was transformed from vortex state to uniform out-of-plane state  by applying the external magnetic field. At the zero field the vortex state is stable and the azimuthal modes with indices $+/-m$ are degenerated. With an increase of the out-of-plane field a continuous transformation to SD state with out-of-plane configuration occurs. In contrast to the ring, low frequency modes, $m=0$ (breathing) and $m=-1$ (gyrotropic) in the soliton state dot are not degenerated due to their coupling to soliton motion. \footnote{In independent simulations we observed the lift of degeneracy for intermediate magnetic states and different pairing of modes at vortex/SD out-of-plane configuration in the ring.} 

\begin{figure}[!ht] 
\includegraphics[width=0.45\textwidth]{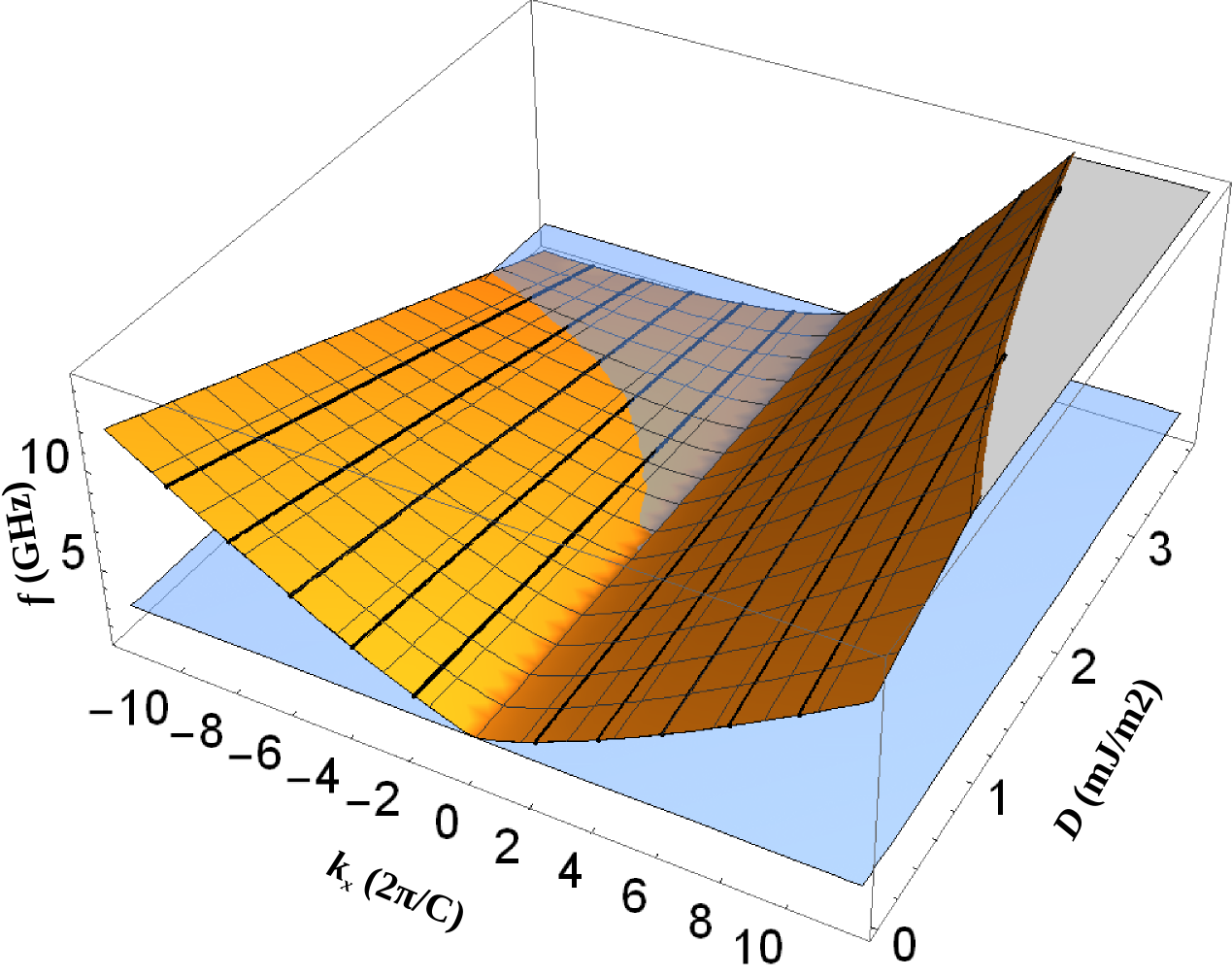} \caption{(Color online) The dispersion relation of spin waves propagating in thin film perpendicular to the in-plane magnetization. Negative group velocity and strong nonreciprocity is pronounced at high DMI parameter values, $D$. Wavevectors corresponding to azimuthal spin waves localized near the skyrmion radius, $R_{\mathrm{s}}=75$ nm, are indicated by the black lines. Several modes are expected to have lower frequency than that of radial $m=0$ ($f_{0}(k=0)$) mode. The frequency of this mode is indicated with opaque blue plane.}
\label{4b} 
\end{figure}

In the ring, the lift of degeneracy for intermediate states and change in the systematics of the SW modes in the vortex and out-of-plane magnetization state was attributed to the geometrical Berry phase present in the vortex state and vanishing in the out-of-plane state.\cite{dugaev2005berry} The same effect explains the lift of degeneracy and change in the systematics of SW frequencies in our investigation. Accounting for the $z$-component of the gauge vector potential \cite{guslienko2016gauge} (resulting in the topological Berry phase) leads to the azimuthal mode index shift from $m$ to $m^{\prime}=m-1$ in the saturated out-of-plane state and degeneracy of the frequencies of the modes having indices $+/-m^{\prime}$ for the edge localized modes in the Bloch skyrmion state. This explains, for instance, the lifting of the degeneracy of the modes $m=+/-2$ increasing $K_{u}$ and apppearing of degeneracy of the azimuthal modes with $m=+4, m^{\prime}=+3$ and $m=-2, m^{\prime}=-3$ in the Bloch skyrmion state (see Fig. 2 and Fig. 3b). The latter degeneracy disappears in course of the transition from the Bloch skyrmion to the N\'eel skyrmion configuration increasing the DMI strength $D$.   

%
%
%


Next, we study the influence of the DMI strength on azimuthal SW modes. With increase of the DMI parameter (path ii $\rightarrow$ iii in Fig. \ref{mathematica} a), the CW and CCW azimuthal modes diverge, a frequency splitting is observed. The frequencies as functions of DMI strength are plotted for the range $D=0$\textemdash $3.8\times10^{-3}$ J/m$^{2}$ in Fig. \ref{Fig4}, where splitting of the azimuthal CW ($m<0$) and CCW modes ($m>0$) is very pronounced at high DMI values.

We can distinguish two causes of the splitting. First, with increase of the DMI strength, size of the skyrmion increases. DMI introduces conditions that favors localization of the azimuthal SW modes at the edge of the skyrmion, see the amplitude localization of dynamical magnetization z-component $\delta m_{z}$ in Fig. \ref{profiles} (c). Edge of the skyrmion serves as a channel for SW propagation, similarly to domain walls in planar nanostructures \cite{winter1961bloch}. The cylindrical symmetry of the static magnetization components is present at the edge of the skyrmion. According to previous observation, the degeneracy should be determined by the number of nodes in the $\delta m_{r}$ and $\delta m_{\phi}$ components. This might affect the SW spectra at low DMI strength and saturate frequency splitting when localization at the skyrmion edge is fully realized.
Second contribution to the frequency splitting of the modes is related to the nonreciprocity of SW induced by DMI \cite{udvardi2009chiral}. Under certain configuration, perpendicular to the in-plane magnetic field, SW possess nonreciprocal properties. Azimuthal SW localized
at the skyrmion edge corresponds to this configuration and the nonreciprocity can influence the azimuthal SWs localized in this region.

Schematic nonreciprocal dispersion of SW in infinite film accounting DMI is presented in Fig. \ref{4b}. We have to note that this model does not allow to estimate SW frequency in the vortex/skyrmion state dot, but shows qualitatively expected effects arising due to the nonreciprocal propagation. To plot this graph we have used the analytical formula for SW propagation in planar film in Damon-Eschbach geometry \cite{kostylev2014interface,di2015asymmetric,stashkevich2015}: 
\begin{widetext}
\begin{equation}
f=\frac{1}{2\pi}\sqrt{(\gamma B_{0}+\omega_{\text{ex}}k^{2})(\gamma B_{0}+\omega_{\text{ex}}k^{2}+\omega_{\mathrm{M}})+\frac{\omega_{M}^{2}}{4}(1-e^{-2|k|t})}+\frac{\gamma Dk}{\pi M_{\mathrm{s}}},\label{eq_1}
\end{equation}
\end{widetext}
where $\omega_{\mathrm{M}}=\gamma\mu_{0}M_{\mathrm{s}}$, $\omega_{\text{ex}}=\frac{2A}{\mu_{0}M_{\mathrm{s}}}$, $B_0$ is a static external magnetic field, and $t$ is the film thickness. The anisotropy is neglected. The DMI introduces the linear term proportional to the wavevector, which decreases the energy of SWs propagating in one direction and increases in the opposite one.

\begin{figure*}
\centering
\includegraphics*[width=0.7\textwidth]{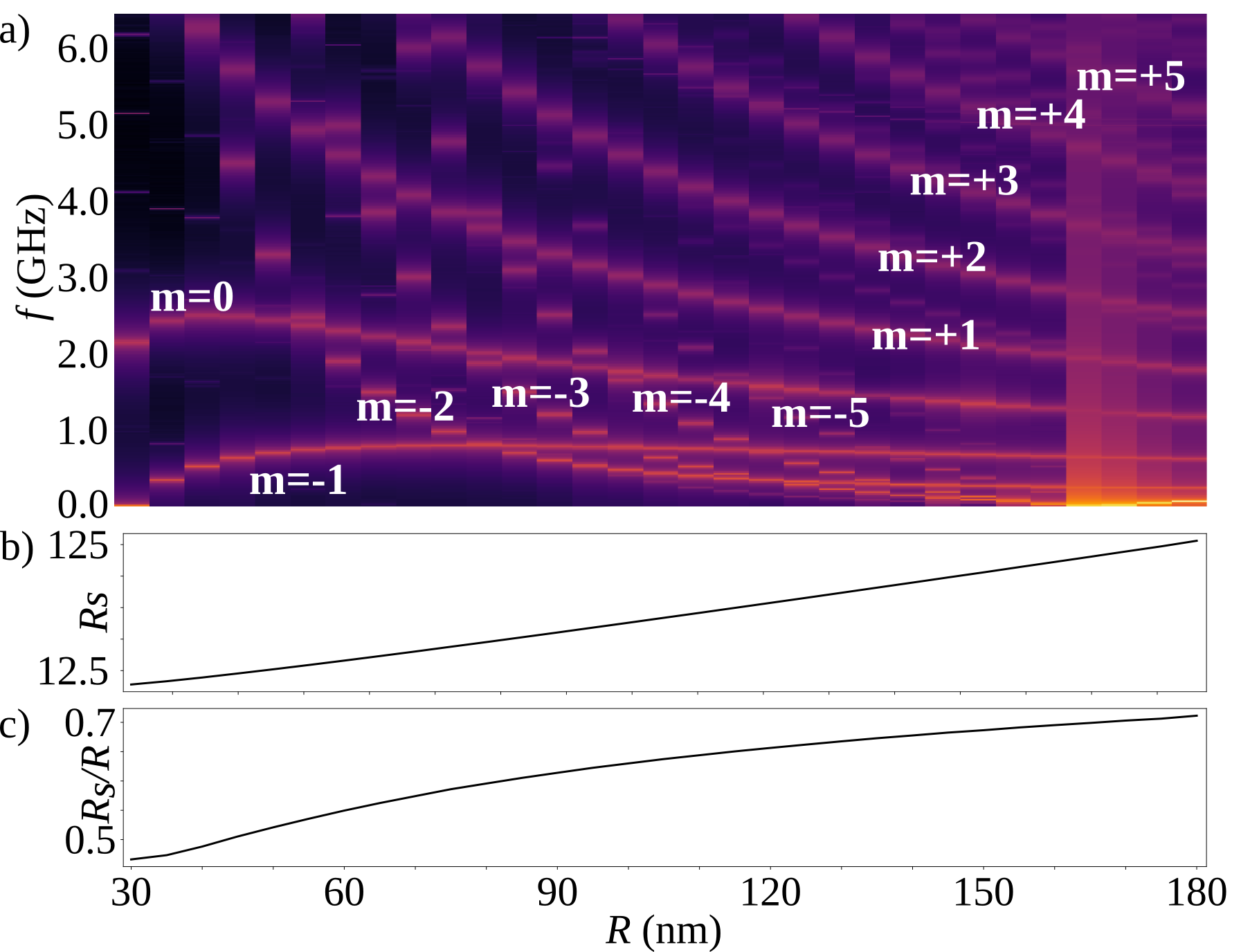} \caption{(Color online) a) Spectrum of the azimuthal spin waves as a function of the dot
radius $R$ at fixed DMI strength, $D=3\times10^{-3}$ J/m$^{2}$. Considerable decrease of
frequency is exhibited for CW propagating spin waves. Coupling of $m=0$ (breathing) and $m=-1$ (gyrotropic)
modes to soliton motion is exhibited as low frequency excitation even for small size skyrmion. b) Skyrmion radius and c) normalized skyrmion radius in dependence on the dot radius.}
\label{Fig5} 
\end{figure*}

\begin{center}
\begin{figure*}
\includegraphics[width=0.7\textwidth]{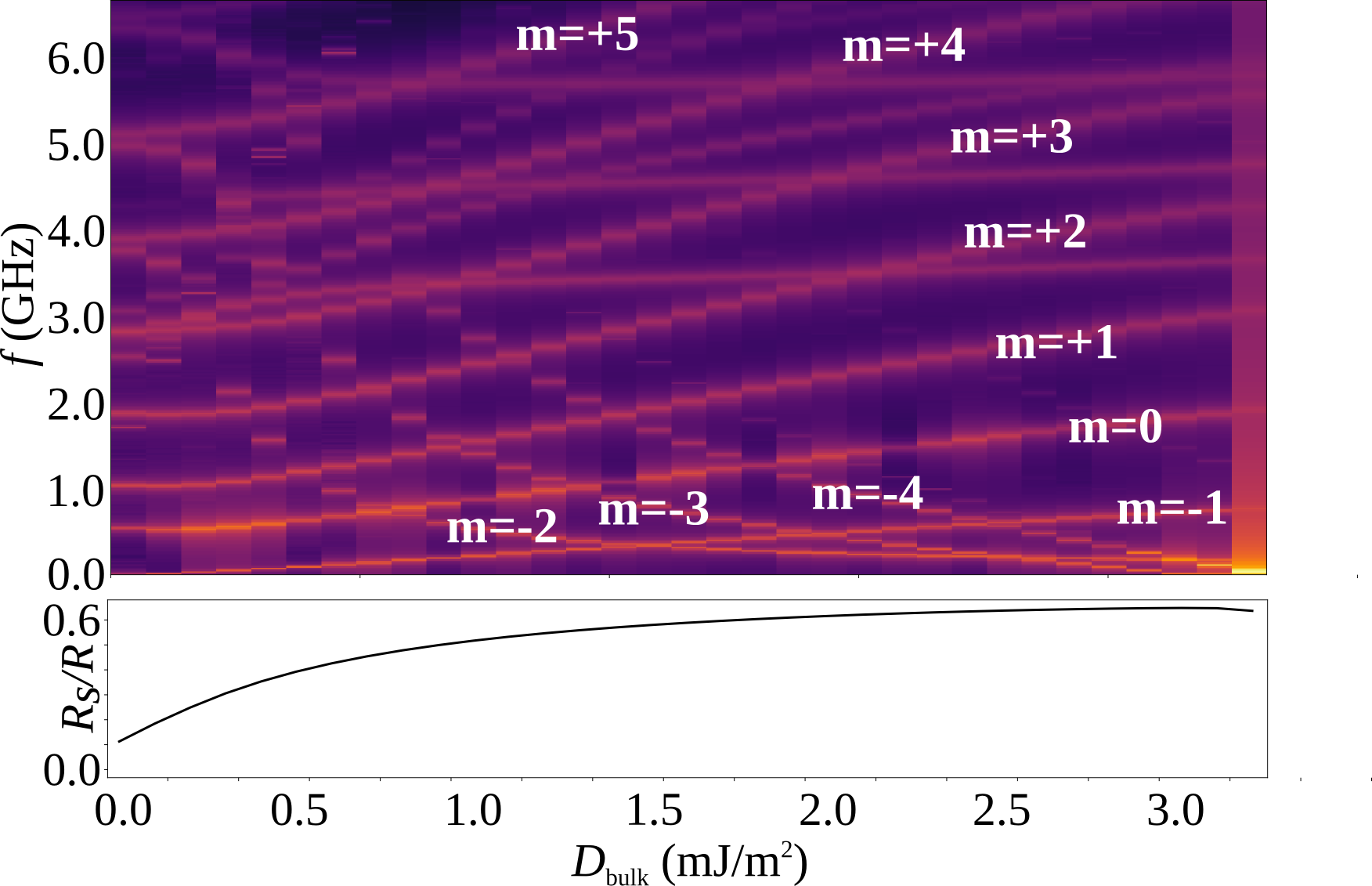} \caption{(Color online) a) Spectrum of the azimuthal modes at fixed value of $K_{\mathrm{u}}=$1.415$\times10^{6}$
J/m$^{3}$ as a function of the bulk DMI strength, $D_{\mathrm{bulk}}$. Splitting of CW and CCW
azimuthal modes is pronounced at high bulk DMI constant, $D_{\mathrm{bulk}}$.
b) Skyrmion size dependence on the DMI strength $D_{\mathrm{bulk}}$. }
\label{Fig4b} 
\end{figure*}
\end{center}

The following parameters were used to approximate the condition of SW propagation at the edge of the skyrmion: $M_{\text{eff}}=0.8$ MA/m \cite{liu2011ferromagnetic}, $B_{0}=0.1$ T, and $C=2 \pi R_{s}$ corresponds to circumference of the skyrmion
with $R_{\mathrm{s}}/R=0.6$. The black lines indicate the wavevectors $k_{m}=m/R_{s}$ which can be related to the azimuthal waves with different $m$ indices ($\lambda=C/m$), blue plane indicates the $f_0=f(k=0)$ frequency, which is related to the mode $m=0$. The figure shows that within the DMI range of 0-3 mJ/m$^{2}$, several $m$ modes with $k<0$ can intersect with $\vert m \vert=0$ mode. At high DMI, close to 3 mJ/m$^{2}$, the SW group velocity is negative for several quantizations, suggesting that the frequency should decrease with quantization of azimuthal SW in the skyrmion state for $m<0$. Indeed, such effect is observed in the dot, see Fig. \ref{Fig4} above $D=2.0$ mJ/m$^{2}$, where crossing of the azimuthal SW mode frequencies is present. The increase of the frequency of modes with the indices $m>0$  is attributed to increase of the frequency of SWs propagating in opposite directions.

The inversion of mode frequencies order for CW waves (decrease of the frequencies with increase of the azimuthal index $\vert m \vert$) is more pronounced with increase of the skyrmion radius, since there is decreasing distance between the wavevectors of these waves. In Fig. \ref{Fig5},
the energy spectrum as a function of the dot radius $R$ is presented in the range of radii 30-180 nm for $D=3$ mJ/m$^{2}$. With increase of the dot radius, the size of the skyrmion increases, as well. The inversion
of the frequency is possible only when the skyrmion size is sufficiently
large. It is interesting to point out here different  behaviour of
the modes $m=0$, and $m=-1$, associated with breathing
and gyrotropic movement of the skyrmion, respectively. Due to the
coupling with soliton motion, the modes exhibit low frequency
even at small skyrmion sizes and the frequency of gyrotropic mode tends
to zero close to the instability of the magnetization configuration. It is another confirmation of different character
of the modes with $m=-1$ and $m=+1$ discussed in the Ref.~[\onlinecite{mruczkiewicz2017spin}].

The nonreciprocity induced by interfacial DMI in SW dynamics in the N\'eel-like skyrmion state 
was shown above. Since bulk DMI introduces the SW frequency nonreciprocity when wave
propagates parallel to in-plane magnetization,\cite{cortes2013influence} thus
similar effects as described above are expected for $D_{\mathrm{bulk}}$ \cite{beg2017dynamics}
that lead to stabilization of a Bloch-like skyrmion. The edge of such skyrmion satisfies the condition for nonreciprocal propagation. Although non-zero $D_{\mathrm{bulk}}$ is not expected in the studied materials, it is instructive to calculate its influence on the SW spectrum. In Fig. \ref{Fig4b} the frequency spectrum
as a function of $D_{\mathrm{bulk}}$ is presented with qualitatively 
the same frequency splitting of the SW modes localized at Bloch-like skyrmion
as for N\'eel-like skyrmion with interfacial DMI, $D$ (Fig.~\ref{Fig4}). This result
can help to understand the frequencies of azimuthal SWs 
observed in Bloch like skyrmions stabilized by $D_{\mathrm{bulk}}$\cite{beg2017dynamics}.

\begin{figure}
\centering
\includegraphics*[width=0.3\textwidth]{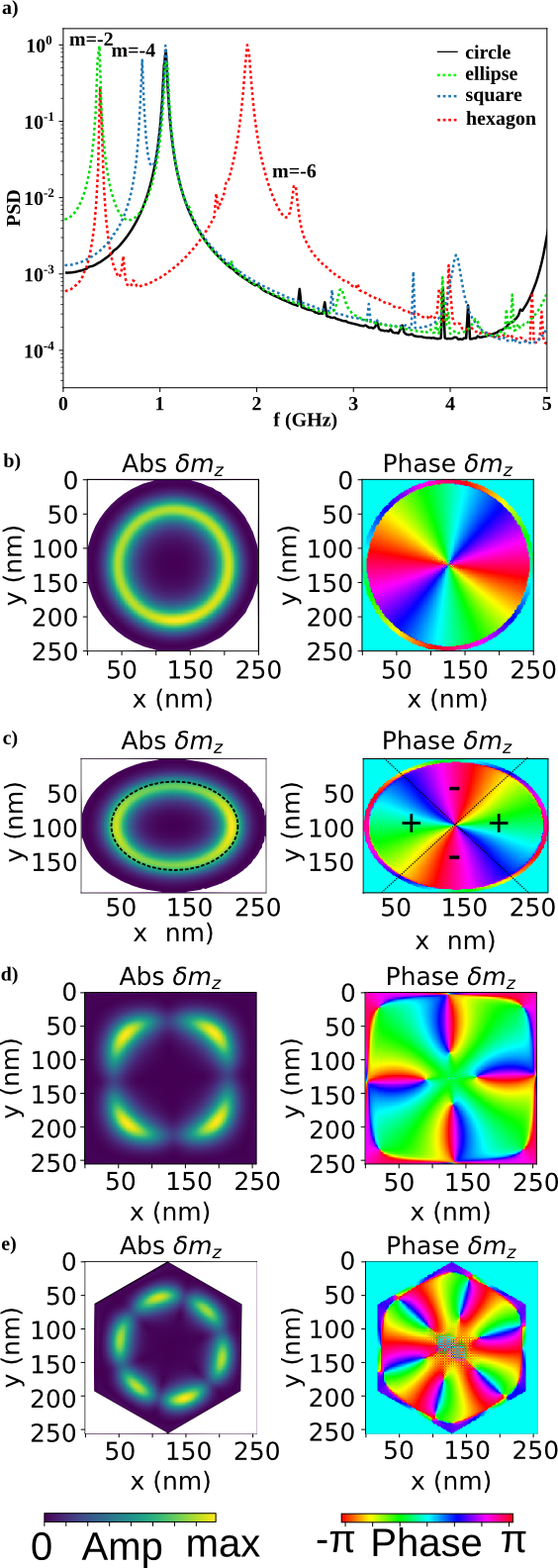} \caption{(Color online) a) PSD spectrum of  circular, elliptical, square and hexagonal shape nanoelements. In (b), (c), and (e) the distribution of the $z$ component of the dynamical magnetization of mode corresponding to symmetry of the nanoelement is presented for circular (radius, $R=125$ nm), square (side length = 250 nm), elliptical (major semi-axis, $a=135$ nm and minor semi-axis $b=115$ nm) and hexagonal (side length, $l=125$ nm) nanoelements, respectively.} 
\label{Fig6} 
\end{figure}

Finally, we study the influence of the dot shape on the coupling efficiency of azimuthal SWs
to the uniform microwave field. Since the skyrmion edge is affected by
the shape of the nanodot, the influence of the shape deformation on the coupling is expected.
As presented in Fig. \ref{Fig6}, for $D=2.2$ mJ/m$^{2}$ ($D=3.4$
mJ/m$^{2}$ for hexagon to lower frequency of $m=-6$ mode)
the mode $m=-2$ has amplitude localized at the edge of
the elliptical skyrmion (indicated as black dashed line). The proportion of the regions with in-phase
and anti-phase is changed, that results in increase coupling to the spatially uniform external magnetic field. Additionally we observe
increase of the amplitude intensity close to the ellipse major semi-axis. This
is not observed for $m=+2$ mode and the coupling is enhanced
only slightly as compared to circular nanodot. We also note that the
influence of elliptical shape on the frequency of the breathing $m=0$ mode
is negligible.

\begin{figure}
\centering
\includegraphics*[width=0.45\textwidth]{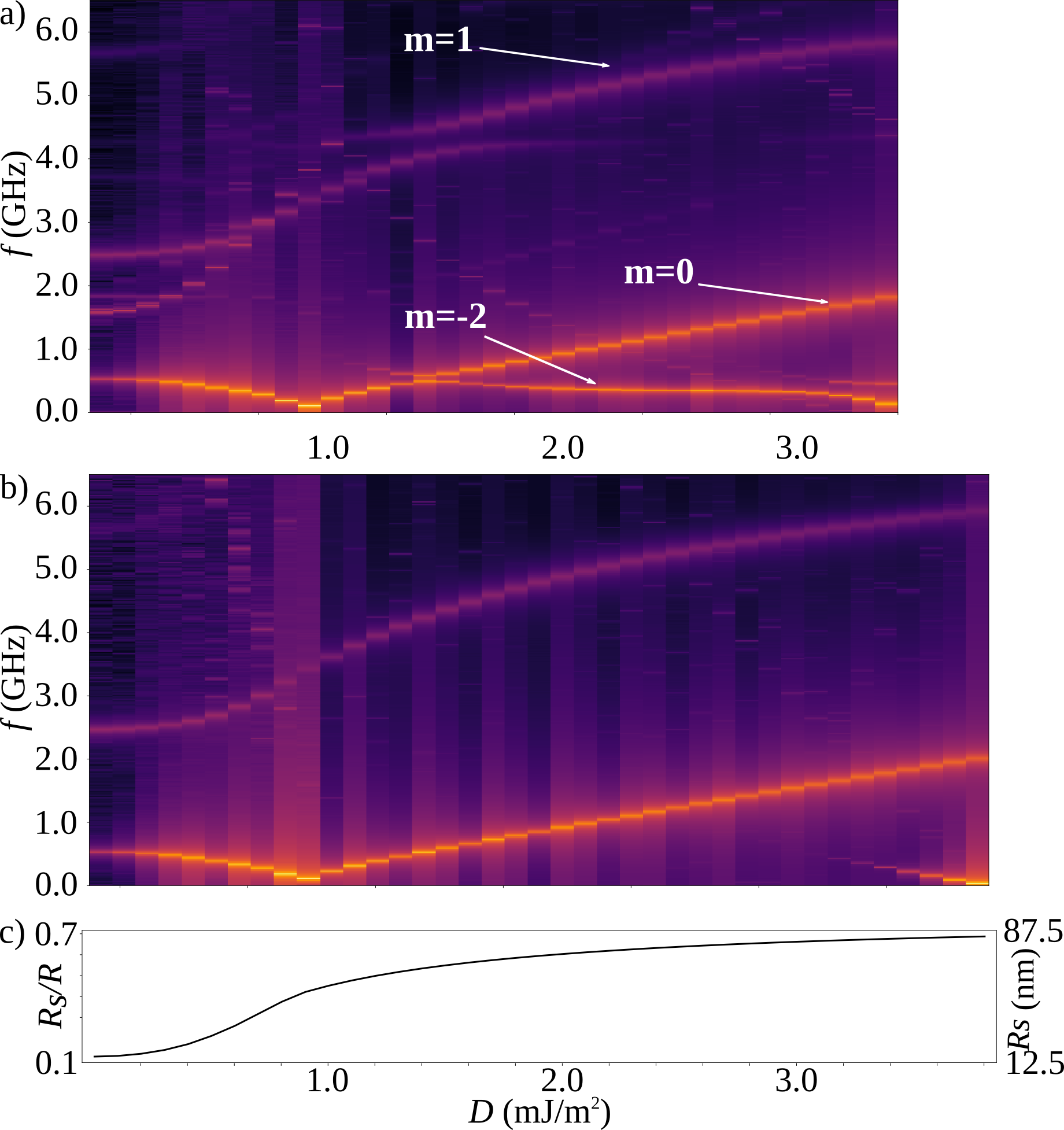} \caption{(Color online) a) Spectrum of SWs in elliptical ($a=135$ nm, $b=115$ nm) nanodot excited by uniform
microwave field (out-of-plane). b) Spectrum of circular nanodot ($R=125$ nm) excited by uniform microwave field (out-of-plane). c) Skyrmion size as a function of the DMI strength $D$ for the circular dot.}
\label{Fig7}
\end{figure}

Similarly, coupling can be relatively strong in nanodots with other
symmetries, like squares and hexagons. We can conclude that the efficiency of coupling to the mode with azimuthal index corresponding to the order of rotational symmetry is decreasing
with increase of the dot symmetry.

The full SW spectrum dependence on $D$ is compared for circular and elliptical nanodots in Fig. \ref{Fig7}. For small $D$, the azimuthal
SW localization is outside of the skyrmion edge. The $m=-2$
mode is not coupled to uniform out-of plane field. Around $D=1$ mJ/m$^{2}$
the skyrmion size increases and the mode localization is shifted to the skyrmion
edge, uniform out-of plane field efficiently excites the $m=-2$
mode. Additionally, a hybridization between the radial quantized mode and azimuthal SW is induced by the symmetry of the ellipse.

\section{Summary\label{Sec:Conclusions}}

We determined the spin excitation spectra of the high order azimuthal SWs in planar nanoelements. The azimuthal SWs found in the vortex state, localized near the edge of the circular dot, are mapped into the azimuthal waves in the Bloch and N'eel skyrmion states. Localization near the edge is present for the vortex state and small size skyrmions, but the degeneracy of mode frequencies is different. The systematics of the frequency levels is different due to geometrical Berry phase which is present for dot-edge localized SWs in the vortex state and absent for the skyrmion states. 

Large size skyrmions show possibility for SW channeling at the skyrmion edge. Due to this localization and presence of DMI, nonreciprocity of azimuthal SWs is relatively strong. The frequency difference of azimuthal SWs propagating in opposite direction is essential and significant mode splitting (between CW and CCW modes) is present. The effect is compared to the DMI induced frequency splitiing of the SWs propagating in an infinite thin film. We showed that the change of nanodot shape (with non-circular symmetry) leads to efficient coupling of the external uniform microwave magnetic field with the azimuthal SW in the skyrmion state having the azimuthal index $\vert m \vert >1$.  
The obtained results allow to understand the skyrmion dynamics in ultrathin nanodots with different shapes and provide a route to manipulate skyrmions in patterned nanostructures, as well as for experimental measurements of the azimuthal SWs. The understanding of the single skyrmion excitations should also help in analysis of magnetization dynamics in the skyrmion lattices.

\section*{Acknowledgements}

The project is financed by the SASPRO Programme. The research received funding from the People Programme (Marie Curie Actions) of the European Union's Seventh Framework Programme under REA grant agreement No. 609427 (Project WEST: 1244/02/01) and was further co-funded by the Slovak Academy of Sciences and the European Union Horizon 2020 Research and Innovation Programme under Marie Sklodowska-Curie grant agreement No.~644348 (MagIC). We acknowledge  financial support of this work by Slovak Grant Agency APVV, grant number APVV-16-0068 (NanoSky) and to Slovak Scientific Grant Agency VEGA, project 2/0183/15. K.G. acknowledges support by IKERBASQUE (the Basque Foundation for Science) and by the Spanish MINECO grant FIS2016-78591-C3-3-R.

\bibliographystyle{apsrev4-1}
\bibliography{books}

\end{document}